\documentclass[aps,prd,twocolumn,preprintnumbers,nofootinbib, amsmath,amssymb,superscriptaddress]{revtex4}

\usepackage{graphicx}
\usepackage{dcolumn}
\usepackage{bm}
\usepackage{amsmath}
\usepackage{simpler-wick}
\usepackage{comment}
\usetikzlibrary{decorations.markings}
\tikzset{W->-/.style={decoration={
  markings,
  mark=at position 0.5*\pgfdecoratedpathlength+2pt with
  {\draw[-latex] (-2pt,0pt) -- (1pt,0pt);}},postaction={decorate}},
  W-<-/.style={decoration={
  markings,
  mark=at position 0.5*\pgfdecoratedpathlength with
  {\draw[latex-] (-2pt,0pt) -- (1pt,0pt);}},postaction={decorate}}
  }
\newif\ifWickBelow
\WickBelowfalse
\pgfkeys{
  /simplerwick/.cd,
  arrows/.store in=\LstWickArrows,
  arrows={-,-,-,-,-,-,-,-,-},
  arrows/.initial={-,-,-,-,-,-,-,-,-}, 
  below/.code={\WickBelowtrue},
}

\makeatletter
\def\swick@end#1#2{
  \swick@setfalse@#1
  \tikzexternaldisable
  \begin{tikzpicture}[remember picture, baseline=(swick-close#1.base)]
    \node[use as bounding box, inner sep=0pt, outer sep=0pt] (swick-close#1) {$\displaystyle #2$};
  \end{tikzpicture}
  \tikz[remember picture, overlay]
{
\foreach \W@X[count=\W@C] in \LstWickArrows
{\ifnum\W@C=#1
\xdef\myW@style{\W@X}
\fi}
\ifWickBelow
    \draw[\myW@style] ($(swick-open#1.south) + (0, -3pt)$) 
          -- ($(swick-open#1.base) + (0, -\swick@offset) + #1*(0, -\swick@sep)$) 
          -- ($(swick-close#1.base) + (0, -\swick@offset) + #1*(0, -\swick@sep)$) 
          -- ($(swick-close#1.south) + (0, -3pt)$);
\else
    \draw[\myW@style] ($(swick-open#1.north) + (0, 3pt)$) 
          -- ($(swick-open#1.base) + (0, \swick@offset) + #1*(0, \swick@sep)$) 
          -- ($(swick-close#1.base) + (0, \swick@offset) + #1*(0, \swick@sep)$) 
          -- ($(swick-close#1.north) + (0, 3pt)$);
\fi}
  \tikzexternalenable}
\makeatother

\newcommand{\beq}{\begin{eqnarray}}
\newcommand{\eeq}{\end{eqnarray}}

\newcommand{\real}{{\sf I}\kern-.12em{\sf R}}
\newcommand{\comp}{{\sf I}\kern-.50em{\sf C}}
\newcommand{\unity}{{\sf I}\kern-.54em{\sf 1}}

\newcommand{\gdll}{\raisebox{-0.4\totalheight}{\includegraphics[scale=.27]{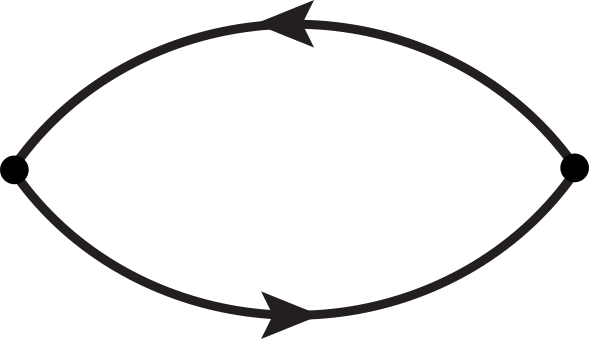}}}
\newcommand{\gdllpm}{\raisebox{-0.4\totalheight}{\includegraphics[scale=.34]{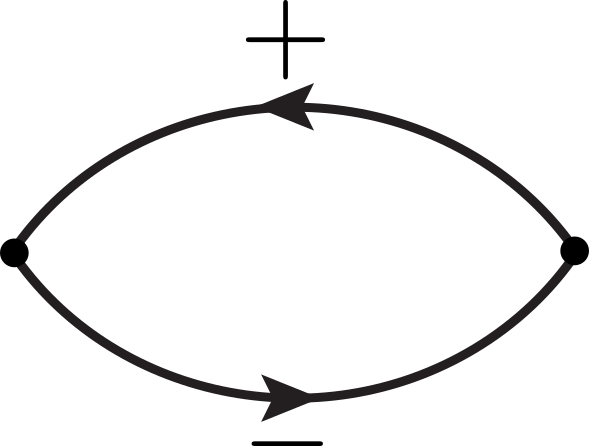}}}
\newcommand{\Mconn}{\raisebox{-0.4\totalheight}{\includegraphics[scale=.27]{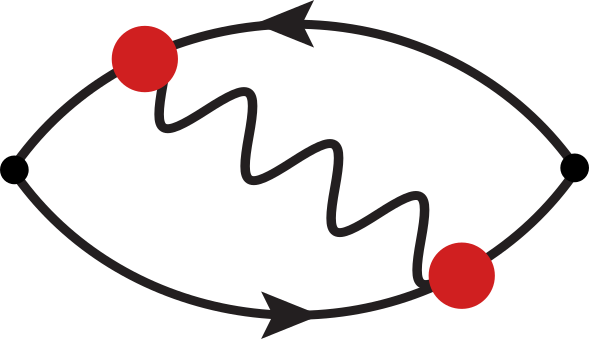}}}
\newcommand{\Mconnpm}{\raisebox{-0.4\totalheight}{\includegraphics[scale=.34]{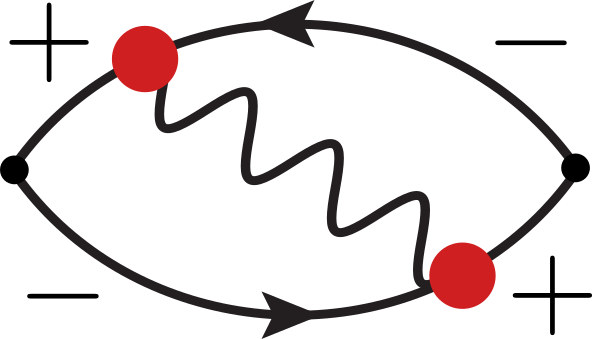}}}
\newcommand{\Mdisc}{\raisebox{-0.2\totalheight}{\includegraphics[scale=.27]{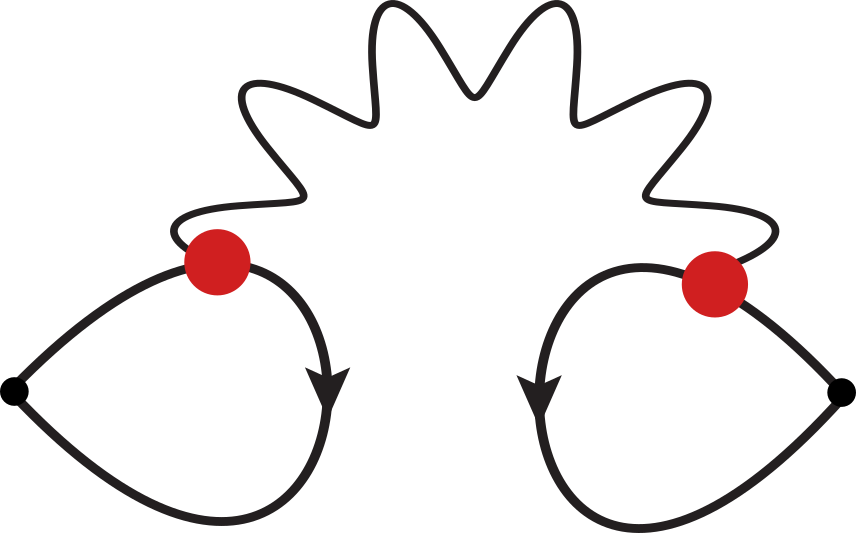}}}
\newcommand{\Mdiscpm}{\raisebox{-0.2\totalheight}{\includegraphics[scale=.34]{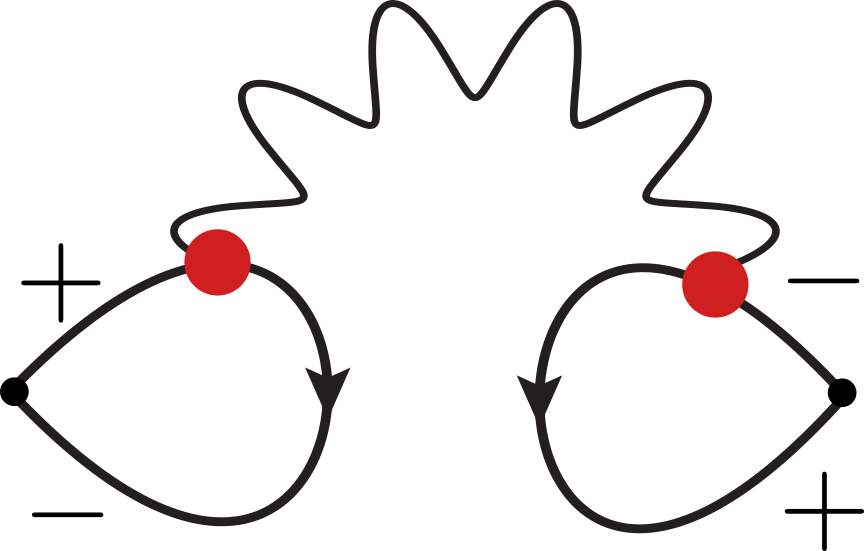}}}
\newcommand{\MconnLarge}{\raisebox{-0.4\totalheight}{\includegraphics[scale=.4]{pics/M3_conn.png}}}
\newcommand{\MdiscLarge}{\raisebox{-0.2\totalheight}{\includegraphics[scale=.4]{pics/M3_disc.png}}}

\newcommand{\tr}{\mbox{Tr}}

\newcommand\defs{\stackrel{def}{=}}

\def\spose#1{\hbox to 0pt{#1\hss}}
\def\ltapprox{\mathrel{\spose{\lower 3pt\hbox{$\mathchar"218$}}
 \raise 2.0pt\hbox{$\mathchar"13C$}}}
 
\usepackage{hyperref}
\hypersetup{
    colorlinks=true,
    linkcolor=blue,
    filecolor=magenta,      
    urlcolor=cyan,
}

\begin{document}

\title{Lattice calculation of the pion mass difference $M_{\pi^{+}}-M_{\pi^{0}}$ at order $\mathcal{O}(\alpha_{em})$ }
\author{R. Frezzotti}
\email{roberto.frezzotti@roma2.infn.it}
\affiliation{Dipartimento di Fisica and INFN, Università di Roma “Tor Vergata”,
Via della Ricerca Scientifica 1, I-00133 Rome, Italy}
\author{G. Gagliardi} 
\email{giuseppe.gagliardi@roma3.infn.it}
\affiliation{Istituto Nazionale di Fisica Nucleare, Sezione di Roma Tre,
Via della Vasca Navale 84, I-00146 Rome, Italy}
\author{V. Lubicz}
\email{vittorio.lubicz@uniroma3.it}
\affiliation{Dipartimento di Fisica, Università Roma Tre and INFN, Sezione di Roma Tre,
Via della Vasca Navale 84, I-00146 Rome, Italy}
\author{G. Martinelli}
\email{guido.martinelli@roma1.infn.it}
\affiliation{Dipartimento di Fisica and INFN Sezione di Roma La Sapienza, Piazzale Aldo Moro 5, I-00185 Rome, Italy}
\author{F. Sanfilippo}
\email{francesco.sanfilippo@infn.it}
\affiliation{Istituto Nazionale di Fisica Nucleare, Sezione di Roma Tre,
Via della Vasca Navale 84, I-00146 Rome, Italy}
\author{S. Simula}
\email{silvano.simula@roma3.infn.it}
\affiliation{Istituto Nazionale di Fisica Nucleare, Sezione di Roma Tre,
Via della Vasca Navale 84, I-00146 Rome, Italy}
\date{\today}

\begin{abstract}
We present a lattice calculation of the charged/neutral pion mass difference $M_{\pi^{+}}-M_{\pi^{0}}$ at order $\mathcal{O}(\alpha_{em})$ using the gauge configurations produced by the Extended Twisted Mass Collaboration with $N_{f}=2+1+1$ dynamical quark flavours at three values of the lattice spacing ($a \simeq 0.062, 0.082, 0.089~{\rm fm}$) and pion masses in the range $M_{\pi} \simeq 250-500~{\rm MeV}$. We employ the RM123 method and expand the path-integral around the isospin symmetric point at leading order in the electromagnetic coupling $\alpha_{em}$. Making use of the recently proposed RTM scheme, we evaluate the full $\mathcal{O}(\alpha_{em})$ contribution, with the inclusion of the disconnected diagram. At the physical point, after performing the continuum and infinite volume extrapolation, we obtain the value $M_{\pi^{+}}-M_{\pi^{0}}=  4.622~(95)~{\rm MeV}$ which is in good agreement with the experimental result $[ M_{\pi^{+}} - M_{\pi^{0}} ]^{exp.} = 4.5936(5)~{\rm MeV}$. 
\end{abstract}

\maketitle

\section{Introduction}
In the last decade, the precision achieved in the computation of several observables relevant for flavour physics by lattice QCD has reached a level where electromagnetic and strong isospin-breaking (IB) effects can no longer be neglected~\cite{Aoki:2021kgd}. The issue of how to evaluate QED effects from lattice QCD simulations has been addressed so far in two different ways: the first method (see e.g. Refs.~\cite{Borsanyi:2014jba,Boyle:2017gzv,Hansen:2018zre}) consists in including QED to the action, performing QCD$+$QED simulations at different values of the electromagnetic coupling $\alpha_{em}$ and extrapolating to the physical value. The second method, the so-called RM123 method~\cite{IBs,deDivitiis:2013xla,lep1,lep2,lep3,gm1,gm2}, consists in expanding the path-integral around the isospin symmetric point $m_{d}=m_{u}$, in powers of the small parameters $\alpha_{em}$ and $m_{d}-m_{u}$, with $\alpha_{em} \sim (m_{d}-m_{u})/\Lambda_{QCD}\sim \mathcal{O}(10^{-2})$. This approach allows to express the expectation value of any given observable in QCD$+$QED as a power series in $\alpha_{em}$ and  $m_{d}-m_{u}$ whose coefficients are related to correlation functions evaluated in the isospin symmetric theory. This has also the advantage that only standard lattice QCD simulations needs to be performed. \\

In the last years, the RM123 method has been successfully applied to the computation of the leading electromagnetic and IB effects to the hadron spectrum, as in the case of the charged/neutral mass splitting of light, stranged and charmed pseudoscalar mesons~\cite{deDivitiis:2013xla,Giusti:2017dmp}. In the case of the neutral pion mass $M_{\pi^{0}}$, its diagrammatic expansion contains a quark-line disconnected diagram connected by a photon line (see Eq.~(\ref{eq:disc_diagram})), in the following simply called ``disconnected diagram'', which contributes to the charged/neutral pion mass difference $M_{\pi^{+}}-M_{\pi^{0}}$. Computing such diagram is a highly non-trival numerical problem, and moreover, it can be shown that, as a consequence of the Dashen theorem~\cite{PhysRev.183.1245}, this disconnected diagram represents a tiny contribution of order $\mathcal{O}(\alpha_{em}\hat{m}_{\ell})$. For this reason it has been neglected in our previous study~\cite{Giusti:2017dmp}. The aim of this paper is to evaluate both the connected and disconnected contributions entering $M_{\pi^{+}}-M_{\pi^{0}}$ at order $\mathcal{O}(\alpha_{em})$.\\

For this calculation, we use the \textit{rotated twisted-mass} (RTM) scheme introduced in Ref.~\cite{Fr:2021}. We have recently shown that this method is particularly convenient for IB and QED calculations based on the RM123 approach, since it allows to consider correlation functions which are affected by much smaller statistical fluctuations with respect to the ones appearing in the RM123 expansion with standard twisted mass (TM) fermions. Evaluating the disconnected diagram appearing in the charged/neutral pion mass difference using the RTM scheme provides in turn an ideal benchmark test in view of its applications to more complicated cases such as the evaluation of the disconnected diagrams relevant for the semileptonic pion and kaon decay ($\pi_{\ell_{3}}$, $K_{\ell_{3}}$). \\

For the numerical simulations, we use the pure QCD isospin symmetric gauge ensembles generated by the Extended Twisted Mass Collaboration (ETMC) with $N_{f}=2+1+1$ dynamical quarks~\cite{Baron:2010bv,Carrasco:2014cwa}. With respect to our previous analysis~\cite{Giusti:2017dmp}, we perform simulations on larger lattice volumes, using two additional ensembles with linear lattice extent $L\sim 3.5$ and $4.2$ fm and lattice spacing $a \sim ~0.089$ fm,  and adopt an improved ansatz in the extrapolation to the physical pion mass and to the continuum and infinite volume limit. Our final result is:
\begin{align}
M_{\pi^{+}}-M_{\pi^{0}} =  4.622~(64)_{stat.}(70)_{syst.}~{\rm MeV}~,
\end{align}
in very good agreement with the experimental value~\cite{pdg}
\begin{align}
\left[ M_{\pi^{+}}-M_{\pi^{0}}\right]^{exp.} = 4.5936(5)~{\rm MeV}~.    
\end{align} \\
The paper is organized as follows: in Sec.~[\ref{sec:lattice_method}] we present our lattice setup and a recap of the RM123 method, focusing on the evaluation of the pion mass splitting in the standard basis and in the RTM scheme. In Sec.~[\ref{sec:numerical_results}] we present our numerical results for $M_{\pi^{+}}-M_{\pi^{0}}$, computed in the RTM scheme with the inclusion of the disconnected diagram. Finally, in Sec.~[\ref{sec:conclusions}] we draw our conclusions.
\section{Methodology}
\label{sec:lattice_method}
\subsection{Lattice discretization of isospin symmetric QCD}
We use the QCD isospin symmetric gauge configurations produced by the ETMC and generated with the maximally twisted Wilson action for fermions, and the Iwasaki action for gluons. All details concerning the lattice discretization have been already presented elsewhere~\cite{Carrasco:2014cwa}. The fermionic action describes $N_{f}=2+1+1$ quark flavors which include in the sea, besides two mass-degenerate quarks, also the
strange and charm quarks with masses close to their physical values.  The untwisted bare quark mass is tuned to its critical value, which guarantees the automatic $\mathcal{O}(a)$ improvement of parity-even observables~\cite{Frezzotti:2003ni}. We consider three values of the inverse bare lattice coupling $\beta$ as well as different lattice sizes. For each lattice spacing, several values of the light sea quark mass are considered in order to perform a reliable extrapolation to its physical value. The complete list of lattice ensembles along with the values of the bare quark masses we use and the number of gauge configurations $N_{cfg}$ accumulated for each ensemble is collected in Tab.~(\ref{tab:simudetails}). \\
\begin{table*}[]
{\small
\begin{center}
\begin{tabular}{||c|c|c|c|c|c|c|c||}
\hline
ensemble & $\beta$ & $a^{-1}~(\rm{GeV})$ & $V / a^4$ & $M_{\pi}~(\rm{MeV})$ & $M_{\pi}L$ &$a\mu_{sea} = a\mu_{val}$ &$N_{cfg}$\\ \hline \hline
$A30.32$ & $1.90$ & $2.227~(85)$  & $32^{3}\times 64$ & $275~(10)$ & $3.95$ &$0.0030$  &$150$ \\
$A40.32$ &  & & & $316~(12)$ & $4.54$ & $0.0040$ & $100$ \\
$A50.32$ &  & & & $350~(13)$ & $5.03$ & $0.0050$ &  $150$ \\
\cline{1-1} \cline{4-8} 
$A40.24$ &  & & $24^{3}\times 48$ & $322~(13)$   & $3.40$ & $0.0040$  & $150$ \\
$A60.24$ &  & &  & $386~(15)$  & $4.18$ & $0.0060$ & $150$ \\
$A80.24$ &  & & & $442~(17)$ & $4.77$ & $0.0080$ &   $150$\\
$A100.24$ & &  & & $495~(19)$ & $5.34$ & $0.0100$ &  $150$ \\
\cline{1-1} \cline{4-8} 
$A40.40$ &  & & $40^{3}\times 80 $ & $317~(12)$ & $5.69$ & $0.0040$ &  $150$ \\
\cline{1-1} \cline{4-8}
$A40.48$ &  & & $48^{3}\times 96 $ & $316~(12)$ & $6.80$ & $0.0040$ & $100$ \\
\hline \hline
$B35.32$ & $1.95$ & $2.418~(84)$  & $32^{3}\times 64$ & $302~(10)$   & $3.98$ &$0.0035$ & $150$  \\
$B55.32$ &   & & & $375~(13)$ & $4.96$ & $0.0055$ & $150$  \\
$B75.32$ &   & & & $436~(15)$ & $5.76$ & $0.0075$ & $~80$ \\
\cline{1-1} \cline{4-8} 
$B85.24$ & & &  $24^{3}\times 48 $ & $468~(16)$ & $4.64$ & $0.0085$ &$150$ \\
\hline \hline
$D20.48$ & $2.10$ & $3.187~(81)$  & $48^{3}\times 96$ & $255~(7)$ & $3.84$ &$0.0020$& $100$ \\ 
$D30.48$ & & & & $318~(8)$  & $4.69$ & $0.0030$ &$100$ \\
 \hline   
\end{tabular}
\end{center}
}
\vspace{-0.25cm}
\caption{\it \small Details of the lattice ensembles we use for the present study. They have been generated by the ETMC employing $N_{f}=2+1+1$ dynamical quark flavors, with degenerate $u$ and $d$ quark masses~\cite{Baron:2010bv, Carrasco:2014cwa}, and correspond to lattice spacings in the range $a \in [ 0.062, 0.089 ]~{\rm fm}$ and pion masses $M_{\pi} \in [250, 500]~{\rm MeV}.$ For each ensemble we also quote the total number $N_{cfg}$ of accumulated independent gauge configurations. }
\label{tab:simudetails}
\end{table*}

\subsection{QED and strong IB corrections: the standard RM123 approach}
\label{sec:RM123}
The leading electromagnetic and strong IB corrections are evaluated adopting the RM123 method~\cite{deDivitiis:2013xla, Giusti:2017dmp}. The starting point is the compact formulation of QED on the lattice in which the photon field $A_{\mu}(x)$ is introduced in the pure QCD lattice action through the replacement
\begin{align}
\label{eq:cov_der_photon}
\nabla_{\mu}\psi_{f}(x) &= U_{\mu}(x)\psi_{f}(x+a\hat{\mu}) -\psi_{f}(x) \nonumber \\[6pt] 
&\to e^{i e q_{f}A_{\mu}(x)}U_{\mu}(x)\psi_{f}(x+a\hat{\mu}) - \psi_{f}(x)~, 
\end{align}
together with the inclusion of the pure gauge photon action $S_{gauge}[A_{\mu}]$ in Feynman gauge
\begin{align}
S_{gauge}\left[A_{\mu}\right] &= \frac{1}{2}{\displaystyle \sum_{x,\mu,\nu}}A_{\mu}(x)\left[ -\nabla_{\nu}^{*}\nabla_{\nu}\right] A_{\mu}(x) \nonumber \\[6pt]
&= \frac{1}{2}{\displaystyle \sum_{k,\mu,\nu}}A^{*}_{\mu}(k)\left[ 2\sin{\left(\frac{k_{\nu}}{2}\right)}\right]^{2}A_{\mu}(k)~.
\end{align}
In Eq.~(\ref{eq:cov_der_photon}), $e^{2}= 4\pi\alpha_{em}$, while $q_{f}$ is the electric charge of the quark $f$ in units of the electric charge of the positron. To cope with the infrared divergence of the photon propagator, we adopt the $QED_{L}$ regularization and set $A_{\mu}(k_{0}, \vec{k}=0)=0$ for all $k_{0}$. The resulting path-integral is then expanded around the isospin symmetric point to the order $\mathcal{O}(\alpha_{em}, \hat{m}_{d} - \hat{m}_{u})$, where $\hat{m}_{u/d}$ are the renormalized masses of the up and down quark in QCD+QED. It should be noted, that the presence of QED interactions produce additional ultraviolet divergences which are then absorbed through a set of properly defined counterterms~\cite{deDivitiis:2013xla}. \\

At leading order, the charged/neutral pion mass splitting is a pure electromagnetic effect, since the leading IB corrections proportional to $m_{d}-m_{u}$ cancel out in both the charged and neutral pion correlators which are symmetric with respect to the exchange $u\leftrightarrow d$. 
According to the analysis of Ref.~\cite{deDivitiis:2013xla}, at order $\mathcal{O}(\alpha_{em})$ the charged/neutral pion mass difference $M_{\pi^{+}} - M_{\pi^{0}}$ is given by
\begin{align}
  \label{eq:pion_splitting_tot_tr}
    M_{\pi^+}- M_{\pi^0} = \frac{e^{2}}{2}(q_{u}-q_{d})^{2}\,\partial_t \,\frac{\delta C_{\pi}^{exch.}(t) - \delta C_{\pi}^{disc.}(t)}{C_{\pi\pi}(t)}~,
\end{align}
where $C_{\pi\pi}(t)$ is the pion correlator of isospin symmetric QCD, while the quantities $\delta C_{\pi}^{exch.}(t)$ and $\delta C_{\pi}^{disc.}(t)$ are defined in terms of the pion correlators of the isospin symmetric theory with two integrated insertions of the electromagnetic current $J_{\mu}(x)$, for which we consider here its local version\footnotemark
\begin{align}
\label{eq:local_em_cur}
J_{\mu}(x) = e~{\displaystyle \sum_{f}}~ q_{f}\bar{\psi}_{f}(x)\gamma_{\mu}\psi_{f}(x)~.    
\end{align}
Explicitly one has\footnotemark
\begin{widetext}
\begin{align}
\label{eq:exch_diagram}
\frac{e^{2}}{2}(q_{u}-q_{d})^{2}\delta C^{exch.}_{\pi}(t) &= -\frac{1}{\sqrt{L^{3}}}{\displaystyle \sum_{\vec{x},y,y'}}\left[\langle 0\big| T \left\{ \wick[arrows={W->-,W-<-, W->-, W->-, W-<-},below]{ \c2 \phi_{\pi^{+}}\c1 (\vec{x},t)~  \c1  J_{\mu}\c4 (y)~ \c2 J_{\nu}\c5 (y')~ \c4 \phi_{\pi^{+}}^{\dag}\c5 (0)}\right\}\big| 0\rangle ~ \Delta_{\mu\nu}(y-y') - \left(\phi_{\pi^{+}}\to \phi_{\pi^{0}}\right)\right] \nonumber \\[7pt]
\implies \delta C_{\pi}^{exch.}(t) &=  \frac{1}{\sqrt{L^{3}}}{\displaystyle \sum_{\vec{x},y,y'}}\langle \tr ~ \bigg[ S_{\ell}((\vec{x},t),y)\gamma_{\mu}S(y,0)\gamma_{5}S_{\ell}(0,y')\gamma_{\nu}S_{\ell}(y', (\vec{x},t))\gamma_{5}      \bigg]  \rangle_{U}~\Delta_{\mu,\nu}(y-y') \nonumber \\[5pt]
&\equiv ~\MconnLarge~, \\[7pt]
\label{eq:disc_diagram}
\frac{e^{2}}{2}(q_{u}-q_{d})^{2}\delta C_{\pi}^{disc.}(t) &= -\frac{1}{\sqrt{L^{3}}}{\displaystyle \sum_{\vec{x},y,y'}}\langle 0\big| T \left\{ \wick[arrows={W->-,W-<-},below]{ \c1\phi_{\pi^{0}}\c2(\vec{x},t) ~\c1 J_{\mu}\c2 (y)~\c1 J_{\nu}\c2 (y')~\c1 \phi_{\pi^{0}}^{\dag}\c2 (0)}\right\}\big| 0\rangle~\Delta_{\mu\nu}(y-y') \nonumber \\[7pt]
\implies \delta C_{\pi}^{disc.}(t) &= \frac{1}{\sqrt{L^{3}}}{\displaystyle \sum_{\vec{x},y,y'}}\langle \tr\bigg[ S_{\ell}((\vec{x},t),y)\gamma_{\mu}S_{\ell}(y, (\vec{x},t))\gamma_{5}\bigg]\cdot \tr\bigg[ S_{\ell}(0,y')\gamma_{\nu}S_{\ell}(y', 0)\gamma_{5}\bigg]  \rangle_{U}~ \Delta_{\mu\nu}(y-y')\nonumber \\[5pt]
&\equiv ~\MdiscLarge ~,
\end{align}
\end{widetext}
where $L$ is the spatial lattice extent,\footnotetext[1]{The use of the (renormalized) local vector current in place of the exactly conserved point-split current adopted in our previous work~\cite{Giusti:2017dmp} gives identical results up to $\mathcal{O}(a^{2})$ lattice artifacts.}\footnotetext[2]{The quark-line connected and disconnected Wick contractions have a relative minus sign stemming from the extra fermion loop present in the disconnected contribution. For convenience, we decided to pull out this extra minus sign from the definition of the disconnected diagram.} $\Delta_{\mu\nu}$ is the photon propagator in Feynman gauge, $\langle 0 | \,.\, | 0\rangle$ is the VEV computed in the isospin symmetric theory, and $\langle \,.\, \rangle_{U}$ indicates the average over the $\textrm{SU}(3)$ gauge field after integrating over the fermionic fields. $S_{\ell}(x,y)$ is the light quark propagator of  isospin symmetric QCD, and the trace $\tr$ is intended over color and Dirac indices. $\phi_{\pi^{+}}$ and $\phi_{\pi^{0}}$ are interpolating operators having the same quantum numbers of the positively charged and neutral pion, for which we choose 
\begin{align}
\phi_{\pi^{+}}(x) &= i\bar{\psi}_{d}(x)\gamma_{5}\psi_{u}(x)~, \\[8pt]
\phi_{\pi^{0}}(x) & = i\frac{\bar{\psi}_{u}(x)\,\gamma_{5}\,\psi_{u}(x) - \bar{\psi}_{d}(x)\,\gamma_{5}\,\psi_{d}(x)}{\sqrt{2}}~.
\end{align} \\

Following the definition given in Ref.~\cite{deDivitiis:2013xla}, the operator $-\partial_{t}$ in Eq.~(\ref{eq:pion_splitting_tot_tr}) corresponds to the evaluation of the so-called effective slope $\delta m_{eff}(t)$ from the ratio of correlators $\delta C/C$,  which is defined through
\begin{align}
\label{eq:def_meff}
\delta m_{eff}(t) &\equiv -\partial_{t} \frac{\delta C(t)}{C(t)} \nonumber \\
&= \frac{1}{F(T/2 -t ,M)}\left(\frac{\delta C(t)}{C(t)} -\frac{\delta C(t-a)}{C(t-a)}\right) ~ ,
\end{align}
where in our case
\begin{align}
\delta C(t) \equiv \delta C_{\pi}^{exch.}(t) - \delta C_{\pi}^{disc}(t),\quad C(t) \equiv C_{\pi\pi}(t)~.
\end{align}
In Eq.~(\ref{eq:def_meff}), $M$ is the ground state mass extracted from the correlator $C(t)$, $T$ is the temporal extent of the lattice and the factor $F(x,M)$ is given by
\begin{align}
F(x,M) &= x\tanh{(Mx)} - (x+a)\tanh{(M(x+a))} ~ .
\end{align}
In the large time limit $t\gg a,~ T-t \gg a$, where the ground state is dominant, the effective slope $\delta m_{eff}(t)$ tends to $M_{\pi^{+}}- M_{\pi^{0}}$. Making use of the previous definitions, we thus simply have 
\begin{align}
\label{eq:pion_diff_final}
M_{\pi^{+}}- M_{\pi^{0}} = \frac{e^{2}}{2}(q_{u}-q_{d})^{2}\partial_{t}\frac{\Mconn - \Mdisc}{\gdll}~,    
\end{align}
where the diagram in the denominator of the r.h.s. is the pion correlator in isospin symmetric QCD. The result of Eq.~(\ref{eq:pion_diff_final}) holds true also in the full unquenched theory. 
As argued in Ref.~\cite{deDivitiis:2013xla}, the disconnected diagram $\delta C_{\pi}^{disc.}$ vanishes in the $\textrm{SU}(2)$ chiral limit as a consequence of the Dashen theorem: in the continuum limit, even in the presence of e.m. interactions, the neutral pion is an exact Goldstone boson for $m_{u}=m_{d}=0$ and arbitrary values of $q_u$ and $q_d$ since the electromagnetic current is diagonal in flavor space and it is invariant under a chiral transformation with generator $\tau_{3}$. This in particular implies (see Ref.~\cite{deDivitiis:2013xla} for details) that the disconnected contribution is of order $\mathcal{O}(\alpha_{em} m_{\ell})$, and represents therefore a tiny corrections which has been neglected in the pioneering analysis of Ref.~\cite{deDivitiis:2013xla} and in the updated study which made use of $N_f=2+1+1$ ensembles of Ref.~\cite{Giusti:2017dmp}. However, aiming to a precise quantitative estimate of the $M_{\pi^{+}}-M_{\pi^{0}}$ mass difference, it is important to evaluate also the disconnected contribution, which is the main motivation for the present work. In addition, the computation of the disconnected diagram entering the pion mass splitting can be considered as a benchmark test in view of the computation of the disconnected diagrams appearing in, e.g., $K_{\ell3}$ decays. 

\subsection{The charged/neutral pion mass difference in the RTM scheme}
\label{sec:isospin_breaking_TM}

In the TM regularization of the lattice fermionic action, isospin is broken at finite lattice spacing already at the level of pure QCD. In particular, the neutral pion correlator involves disconnected diagrams, which cancel in the continuum due to isospin symmetry. Moreover, the related mass suffers from $\mathcal{O}(a^{2}\Lambda_{QCD})$ cut off effects. On the contrary, the charged pion correlator involves only connected diagram, and its mass is affected only by $\mathcal{O}(a^{2}m_l)$ cut-off effects, behaving as a true Goldstone boson as shown in Refs.~\cite{Frezzotti:2003ni, Frezzotti:2004wz}. In addition, correlation functions involving the propagation of the TM neutral pion, are typically noisier than the corresponding charged ones. Therefore, the disconnected contribution to $M_{\pi^{+}}-M_{\pi^{0}}$ is very difficult to determine in the standard TM approach, due to the presence of  large statistical fluctuations in the neutral sector.\\

In our previous analysis of the pion mass splitting~\cite{deDivitiis:2013xla,Giusti:2017dmp}, we evaluated Eq.~(\ref{eq:exch_diagram}) in a mixed action setup. In the valence sector, for each light quark flavor $f$, a pair of Osterwalder-Seiler fermions $\psi_{f}^{+}, \psi_{f}^{-}$, having the same mass of the corresponding sea quark, but regularized with opposite values of the Wilson parameter, $r=\pm 1$, has been introduced. This allowed to compute the correlation functions in Eq.~(\ref{eq:exch_diagram}) using as interpolating field of the neutral and charged pion: \begin{align}
\label{eq:pi_plus_mixed}
\phi_{\pi^{+}}(x) & = i\bar{\psi}_{d}^{-}(x)\,\gamma_{5}\,\psi_{u}^{+}(x)~, \\[8pt]
\label{eq:pi_zero_mixed}
\phi_{\pi^{0}}(x) & = i\frac{\bar{\psi}_{u}^{+}(x)\,\gamma_{5}\,\psi_{u}^{-}(x) - \bar{\psi}_{d}^{+}(x)\,\gamma_{5}\,\psi_{d}^{-}(x)}{\sqrt{2}}~.   
\end{align}
 In this way both charged and neutral pions contain valence quarks carrying different signs of the twisted Wilson term. This choice gives rise to correlators having reduced statistical errors compared to other choices of the interpolator fields like $\bar{\psi}_{f}^{+}\psi_{f}^{+}$. However, within this mixed action setup, only the connected fermionic Wick contraction in Eq.~(\ref{eq:exch_diagram}) is non-zero, and in the continuum limit  the corresponding correlation function only reproduces the correct contribution of $\delta C_{\pi}^{exch}$ to the pion mass splitting. In Refs.~\cite{deDivitiis:2013xla, Giusti:2017dmp}, the more noisy contribution $\delta C_{\pi}^{disc.}(t)$ has been neglected. \\

In a recent paper~\cite{Fr:2021}, we showed that it is possible to evaluate both the exchange and disconnected diagrams by considering only correlation functions involving quark lines with opposite values of the Wilson parameter by working in the so-called RTM scheme. In this scheme, the pure isoQCD action of the light-quark sector is given by~\cite{Fr:2021}
\begin{align}
\label{eq:RTM}
\mathcal{L}_{RTM}(\psi'_{\ell}) &= \bar{\psi}'_{\ell}(x)\left[ \gamma_{\mu}\tilde{\nabla}_{\mu}  -i\gamma_{5}\tau_{3}W(m^{cr.})+ \hat{m}_{\ell}\right]\psi'_{\ell}(x)~,  
\end{align}
where $\psi'_{\ell}= (u',d')$, while $\widetilde \nabla_\mu$ is the lattice symmetric covariant derivative, written in terms of the forward $(\nabla_\mu )$ and backward ($ \nabla^*_\mu$) covariant derivatives, 
\begin{equation}
\widetilde{\nabla}_\mu = \frac{1}{2} \left( \nabla^*_\mu + \nabla_\mu \right)~.
\end{equation}
In Eq.~(\ref{eq:RTM}), $W(m_{cr})$ is the critical Wilson term, which includes the critical mass $m_{cr}$, and it is given by
\begin{equation}
W(m_{cr}) = - a\, \frac{r}{2}\, \nabla_\mu \nabla^*_\mu + m_{cr}(r)\, .
\end{equation}
In the RTM scheme, the primed quark fields $u',d'$ are regularized with opposite values of the Wilson parameter $r=\pm 1$, and are related to the ``physical'' basis up and down quark fields, $u, d$, through the rotation
\begin{align}
\label{eq:rotation_RTM}
\begin{pmatrix}
u' \\
d' 
\end{pmatrix}
= \frac{1}{\sqrt{2}}\begin{pmatrix}
1 & 1 \\
-1 & 1
\end{pmatrix}
\begin{pmatrix}
u \\
d 
\end{pmatrix}~.
\end{align}
Notice that the RTM action is not equivalent to the standard TM action, because the transformation in Eq.~(\ref{eq:rotation_RTM}) does not correspond to a symmetry of the discretized theory due to the presence of the twisted Wilson term proportional to $\tau_{3}$. We call here ``physical'' quark basis the one where the isospin breaking terms proportional 
to $(m_u-m_d)/2 $ and $(q_u-q_d)/2 = \Delta q$, once written in terms of $\psi_\ell = (u,d)^T$,
involve a $\tau_3$ matrix. In the primed basis of Eq.~(\ref{eq:RTM}) these isospin violating  terms, 
expressed via the fields $\psi' _\ell= (u',d')^T$, involve instead a $\tau_1$ matrix. 
The peculiarity of the RTM valence fermion action is that different Pauli matrices (e.g.\ 
$\tau_3$ and $\tau_1$) appear in the chirally twisted Wilson term and in the isospin breaking 
terms, whatever basis is taken.\\

In the basis of Eq.~(\ref{eq:RTM}), the ``rotated charged'' pion fields $\phi_{\pi'^{-}}= \bar{u}'\gamma_{5}d'$ and $\phi_{\pi'^{+}}= \bar{d}'\gamma_{5}u'$ are related to the physical pion fields through
\begin{align}
\label{eq:rot_field_transf}
\phi_{\pi'^{-}} &= -\frac{1}{2}\left( \phi_{\pi^{+}} - \phi_{\pi^{-}}\right) - \frac{1}{\sqrt{2}}\phi_{\pi^{0}}~, \nonumber \\
\phi_{\pi'^{+}} &= \frac{1}{2}\left( \phi_{\pi^{+}} - \phi_{\pi^{-}}\right) - \frac{1}{\sqrt{2}}\phi_{\pi^{0}}~,
\end{align} 
while the light quark contribution to the electromagnetic current in Eq.~(\ref{eq:local_em_cur}), written in the rotated basis, takes the form
\begin{align}
J_{\mu}(x)  \to J'_{\mu}(x) \equiv  \bar{J}_{\mu}(x) + J_{\mu}^{ib}(x)~,     
\end{align}
where
\begin{align}
\label{eq:RTM_em_cur}
\bar{J}_{\mu}(x) &\equiv e\bar{q}\,\bar{\psi}'_{\ell}(x)\gamma_{\mu}\psi'_{\ell}(x) \\[8pt]
J_{\mu}^{ib}(x) &\equiv  -e\Delta q\,\bar{\psi}'_{\ell}(x)\tau_{1}\gamma_{\mu}\psi'_{\ell}(x)
\end{align}
with $\bar{q}= (q_{u}+q_{d})/2$ and $\Delta q=(q_{u}-q_{d})/2$. The IB component of the electromagnetic current $J_{\mu}^{ib}$ in Eq.~(\ref{eq:RTM_em_cur}) has an unconventional direction in flavor space, and induces a mixing between the $u'$ and $d'$ quarks. \\

Using Eq.~(\ref{eq:rot_field_transf}) it is easily realized that the correlator $C_{\pi'^{+}\pi'^{-}}(t) = \langle 0| \phi_{\pi'^{+}}(t) \phi^{\dag}_{\pi'^{-}}(0)| 0 \rangle$ which describes, in the rotated basis, the mixing between the positive and negative rotated pion fields, is related to the difference between the charged and neutral physical pion correlator via
\begin{align}
C_{\pi'^{+}\pi'^{-}}(t) = \frac{1}{2}\left(\,C_{\pi^{0}\pi^{0}}(t) - C_{\pi^{+}\pi^{+}}(t)\,\right)~.
\end{align}
The mixing is absent in the QCD isospin symmetric theory, and it is generated, in the RTM scheme, by a double insertion of the ``flavour off-diagonal'' component $J_{\mu}^{ib}$ of the rotated electromagnetic current, which is an $\mathcal{O}(\alpha_{em})$ effect. From this result it follows (see Ref.~\cite{Fr:2021} for a detailed derivation), that the pion mass difference at $\mathcal{O}(\alpha_{em})$ can be expressed in this scheme as \vspace{0.25cm}
\begin{align}
\label{corr_RTM}
M_{\pi^{+}} - M_{\pi^{0}} = ~&2e^{2}(\Delta q)^{2}Z_{A}^{2}~\times \nonumber  \\[10pt]
&\partial_{t}\frac{\overbrace{\Mconnpm}^{\textstyle \delta C'^{exch.}_{\pi}(t)}\,\, - \,\,\overbrace{\Mdiscpm}^{\textstyle \delta C'^{disc.}_{\pi}(t)}}{\underbrace{\gdllpm}_{\textstyle C_{\pi\pi}^{\rm{isoQCD}}(t)}}~,
\end{align}
where we now also include explicitly the RC $Z_{A}$ of the local current $J_{\mu}^{ib}$, which in our TM setup renormalizes as the axial current for untwisted Wilson quarks. In the diagrams of Eq.~(\ref{corr_RTM}), we show explicitly the sign of the Wilson parameter on each quark line, which gets always flipped at the e.m. vertex where the $u^{\prime}$ quark turns into a $d^{\prime}$ quark and viceversa. Manifestly, in both exchange and disconnected diagrams, only the ``rotated charged'' isospin symmetric pion propagates, and the resulting correlation functions are affected by strongly reduced statistical errors. We refer to Ref.~\cite{Fr:2021} for a statistical comparison between rotated and unrotated disconnected correlators as well as for a more detailed discussion about the renormalization properties and continuum limit of the correlation functions in Eq.~(\ref{corr_RTM}). In the next section, we evaluate the charged/neutral pion mass splitting from the diagrams displayed in Eq.~(\ref{corr_RTM}).

\section{Numerical Results}
\label{sec:numerical_results}
\begin{figure*}
    \centering
    \includegraphics[scale=0.7]{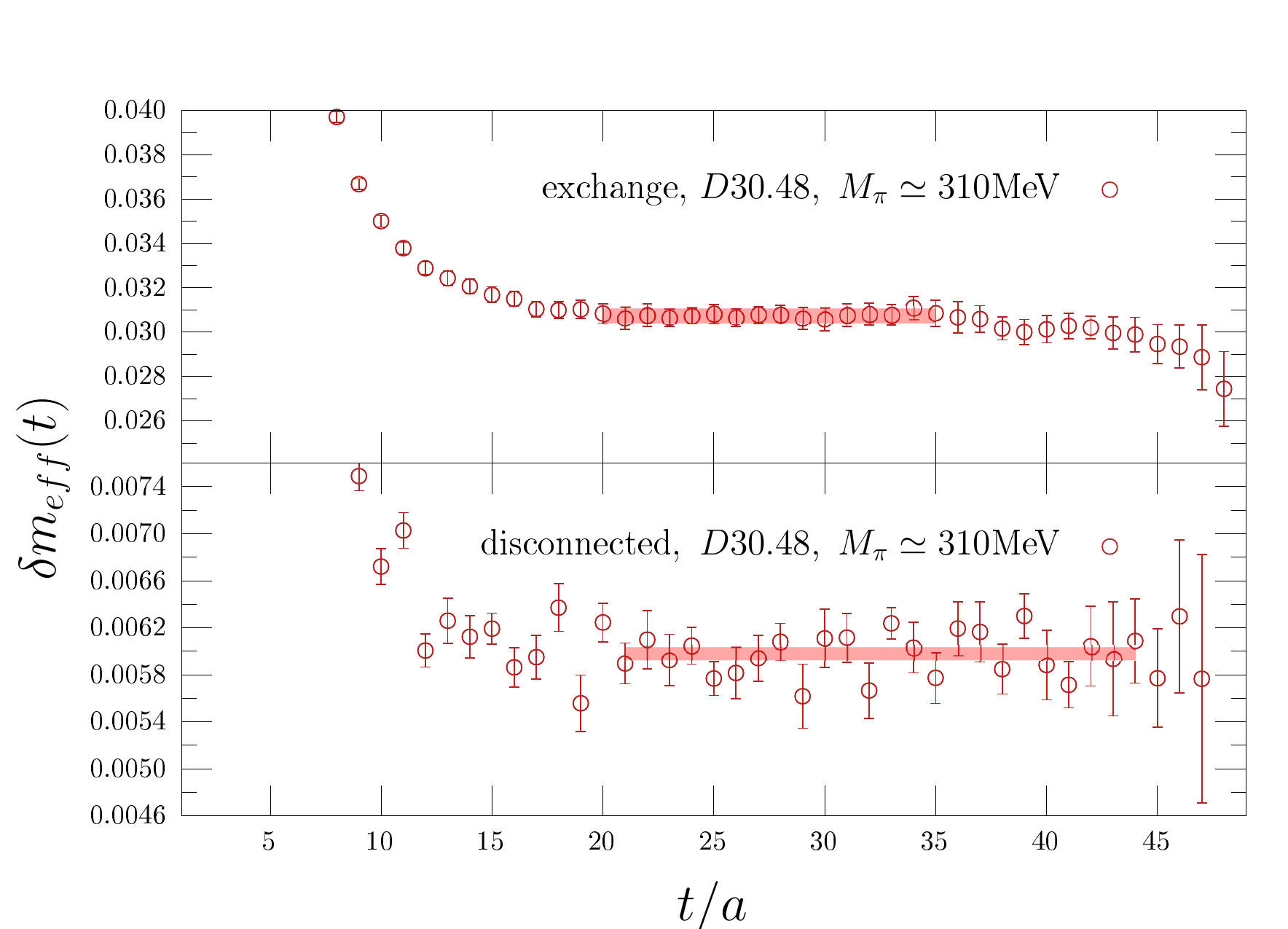}
    \caption{\small\it Time dependence of the effective slope $\delta m_{eff}(t)$, in the case of the exchange (top) and of the disconnected (bottom) diagram, for the gauge ensemble $D30.48$. The horizontal bands indicate the result of a constant fit in the plateaux region  where the ground state dominates. }
    \label{fig:example_plateaux}
\end{figure*}

As already mentioned, for this study we use the  $N_{f}=2+1+1$ ensembles of Wilson TM fermions generated by the ETMC by only considering the ensembles with $M_{\pi}L > 3.8$. With respect to Tab.~(\ref{tab:simudetails}), we thus exclude from the final analysis the ensemble $A40.24$. This subset of ensembles correspond to pion masses in the range $M_{\pi} \in [250, 500]~{\rm MeV}$ and lattice spacings from $a\sim 0.089~{\rm fm}$ down to $a\sim 0.062~{\rm fm}$. For the RC $Z_{A}$ appearing in Eq.~(\ref{corr_RTM}), we use the precise determination obtained from the method $M_{2}$ of Ref.~\cite{Carrasco:2014cwa}, namely $Z_{A} = \{ 0.703(2), 0.714(2), 0.752(2)\}$ at $\beta =\{ 1.90, 1.95, 2.10\}$.
To improve the precision on the disconnected diagram of Eq.~(\ref{corr_RTM}), we devised a new numerical technique, tailored for quark disconnected diagrams, in which the photon propagator is evaluated exactly by working in momentum space, and therefore the statistical noise generated by its stochastic representation is absent. The method, which is discussed in details in App.[\ref{sec:num_methods}], combined with the benefit of the RTM scheme, allows us to obtain an $\mathcal{O}(1\%)$ statistical accuracy on the value of the disconnected diagram. As an example, we show in Fig.~(\ref{fig:example_plateaux}), for the ensemble D30.48 (see Tab.~(\ref{tab:simudetails})), our determination of the effective slope $\delta m_{eff}(t)$ extracted from both the exchange and the disconnected diagram. \\    

To reduce the sensitivity of the result to the uncertainties of the scale setting, we find it useful to consider the dimensionless ratio
\begin{align}
   R_{\pi} \equiv \frac{M^{2}_{\pi^+} -M^{2}_{\pi^0}}{f_{\pi}^{2}} \approx  \frac{2M_{\pi}}{f_{\pi}^{2}}\left( M_{\pi^{+}} - M_{\pi^{0}}\right)~,
\end{align}
where $f_{\pi}$ is the pion decay constant, and $M_{\pi}$ is the mass of the TM charged pion in isospin symmetric QCD. Aiming at a determination of the pion mass splitting with $\mathcal{O}(1\%)$ accuracy, it is however important to have control over the QCD exponentially suppressed FSEs  affecting both $f_{\pi}$ and $M_{\pi}$, to which we apply the $\rm{SU}(2)$ ChPT finite volume corrections at NNLO $+$ resummation, i.e. the Colangelo-D{\"u}rr-Haefeli (CDH) formulae~\cite{Colangelo:2005gd}. The latter depend on the knowledge of the four, scale dependent, $\rm{SU}(2)$ low-energy constants (LECs) $\bar{\ell}_{1}, \bar{\ell}_{2}, \bar{\ell}_{3}, \bar{\ell}_{4}$, for which in this work we adopt the values $\bar{\ell}_{1}(M_{\pi}^{phys.}) = -0.4$, $\bar{\ell}_{2}(M_{\pi}^{phys.}) = 4.3$, $\bar{\ell}_{3}(M_{\pi}^{phys.}) = 3.2$, $\bar{\ell}_{4}(M_{\pi}^{phys.}) = 4.4$. The size of such corrections is in all cases smaller than one percent. \\

\begin{figure*}
    \centering
    \includegraphics[scale=0.65]{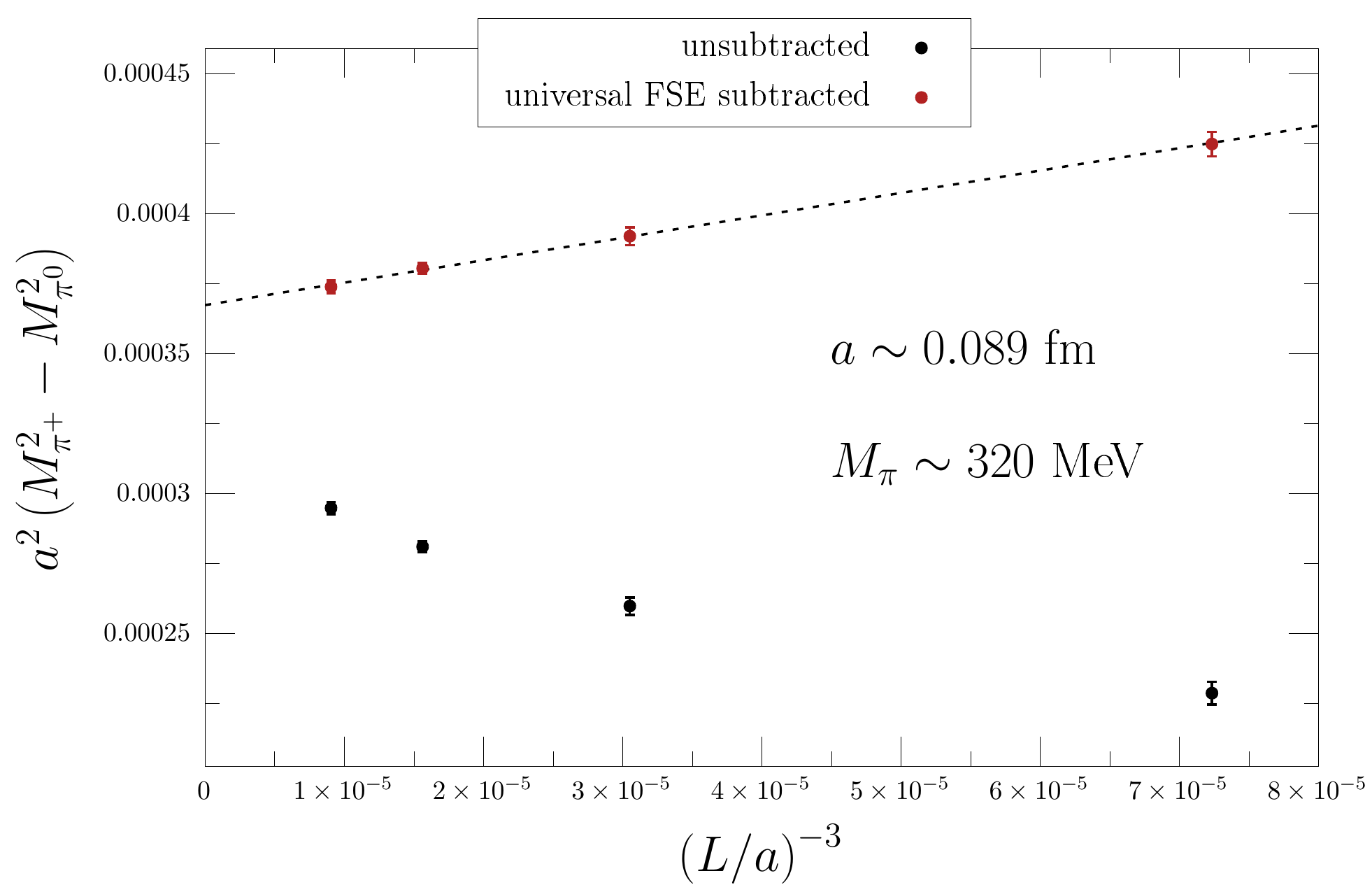}
    \caption{\small\it The squared pion mass difference in the RTM scheme, for the ensembles of type $A40.XX$, which share a common value of the pion mass ($M_{\pi}\simeq 320~\textrm{MeV}$) and of the lattice spacing, but differ in the lattice size $L$. The black points represent the data in lattice units without any FSE correction, while the red ones are corrected subtracting the universal FSEs of Eq.~(\ref{eq:universal_FSE}). The dashed line is the result of a linear fit.         }
    \label{fig:A40}
\end{figure*}

When a massless photon is put on a box, FSEs show up in QED  observables as inverse powers of the spatial extent. The analysis of such finite volume corrections has been the subject of several studies (see Refs.~\cite{Davoudi:2014qua, Borsanyi:2014jba}), where different infrared regularizations of QED have been considered. FSEs on hadron masses start at order $\mathcal{O}(L^{-1})$ and they are universal up to order $\mathcal{O}(L^{-2})$ included, i.e. their size depends solely on the charge, mass and spin of the hadron, but not on its internal structure. In the case of the $QED_L$ which we use in this work, and for a pseudo-scalar meson of electric charge $Q$ and mass $M_{PS}$, the universal FSEs are given by
\begin{equation}
M_{PS}^2(L) - M_{PS}^2(\infty) = - Q^2 \alpha_{em} \frac{2\kappa}{L^2}\left( 1 +  \frac{1}{2}M_{PS} L \right) ~ ,
     \label{eq:universal_FSE}
\end{equation}
where $\kappa=2.837297$. Such corrections have been applied to our lattice data leaving residual structure-dependent (SD) $\mathcal{O}((1/L)^{3})$ FSEs. This is shown in Fig.~(\ref{fig:A40}) for the four ensembles of type $A40.XX$, which only differ in the spatial extent. For these ensembles, the residual FSEs are very well described by a $1/L^{3}$ term, and we do not see evidence of higher order $\mathcal{O}(1/L^{4})$ FSEs within errors.   \\

The SD FSEs have been analyzed in Ref.~\cite{Davoudi:2014qua} on the basis of a non-relativistic expansion.  The leading $\mathcal{O}(L^{-3})$ term is found to be proportional to the squared pion charge radius $\langle r^2 \rangle_{\pi^+}$ via
\begin{align}
 \left[ M_{\pi^+}^2(L) - M_{\pi^+}^2(\infty) \right]^{(SD)} =\frac{e^2}{3} \frac{M_\pi}{L^3} \langle r^2 \rangle_{\pi^+} +
                                                                                       {\cal{O}}(\frac{1}{L^4}, \frac{M_\pi}{L^4}),
    \label{eq:pion_savage}
\end{align}
where $\langle r^2\rangle_{\pi^+} = (0.672 \pm 0.008 ~ \mbox{fm})^2$. At order $\mathcal{O}(L^{-4})$ QED-related FSEs starts to be present also for neutral pseudo-scalar particles.  When extrapolating towards the thermodynamic limit, we make use of Eq.~(\ref{eq:pion_savage}) to remove the SD FSEs, taking into account possible deviations from the non-relativistic prediction. \\

\begin{figure*}
    \centering
    \includegraphics[scale=0.74]{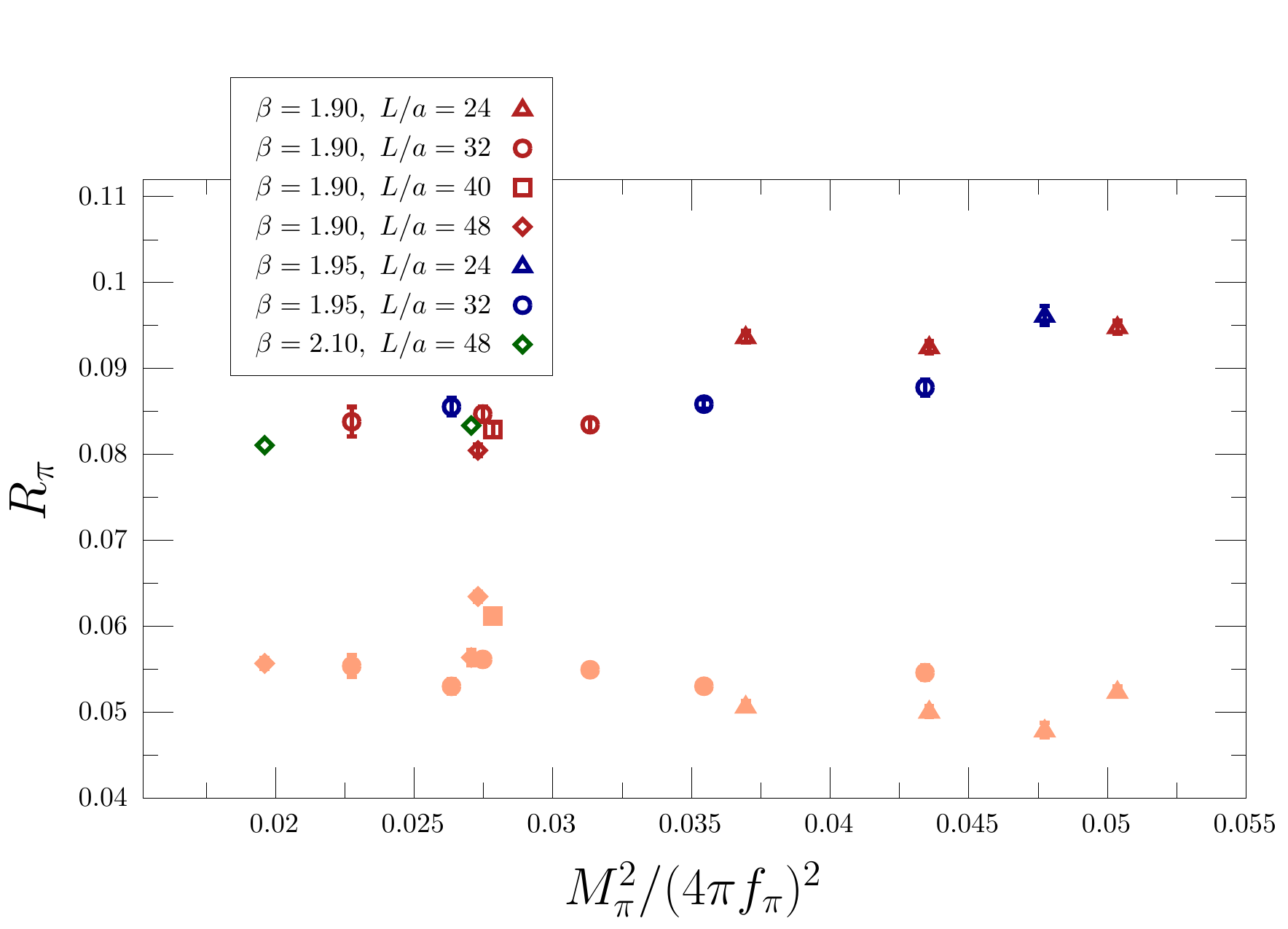}
    \caption{\small\it Our results for the ratio $R_{\pi}$ as a function of the dimensionless ratio $M_{\pi}^{2}/(4\pi f_{\pi})^{2}$, determined in the RTM scheme and including the contribution of the disconnected diagram. The filled markers represent the data without any FSE correction, while the empty ones represent the result of the subtraction of the universal FSE using Eq.~(\ref{eq:universal_FSE}). }
    \label{fig:univ_and_raw_data}
\end{figure*}

In Fig.~(\ref{fig:univ_and_raw_data}) we show our determination of the $R_{\pi}$ ratio, before and after removal of the universal FSEs of Eq.~(\ref{eq:universal_FSE}), as a function of the dimensionless ratio $M_{\pi}^{2}/(4\pi f_{\pi})^{2}$. As the figure shows, the statistical accuracy is very good and in all cases of order of percent or smaller. The universal FSEs corrections are always sizable on our lattice volumes, approaching $45\%$ at the largest simulated value of $M_{\pi}^{2}/(4\pi f_{\pi})^{2}$. \\

According to the analysis of Ref.~\cite{Hayakawa:2008an}, the $\rm{SU}(3)$ ChPT prediction for the squared pion mass difference at NLO is given by
\begin{align}
&M_{\pi^+}^{2} - M_{\pi^0}^{2} = e^{2}f_{0}^{2}C\left[ 1 - 4\left(\frac{M_{\pi}}{4\pi f_{0}}\right)^{2}\log{\left(\frac{M_{\pi}}{4\pi f_{0}}\right)^{2} } \right. \nonumber \\[6pt]
& \left. -2\left(\frac{M_{K}}{4\pi f_{0}}\right)^{2}\log{ \left(\frac{M_{K}}{4\pi f_{0}}\right)^{2}}\right]  \nonumber \\[6pt]
& -3e^{2}f_{0}^{2}\left(\frac{M_{\pi}}{4\pi f_{0}}\right)^{2}\log{\left(\frac{M_{\pi}}{4\pi f_{0}}\right)^{2} } + \mathcal{O}\left(M_{\pi}^{2}, M_{K}^{2}\right)~,
\end{align}
where $f_{0}$ is the pion decay constant in the chiral limit, and $C$ is a LEC. Here and in the following two equations, we focus on the terms
with the leading logarithmic dependence on the pseudoscalar light meson masses and denote by ${\mathcal O}(M_\pi^2,M_K^2)$ terms of first and higher order in the squared pion and kaon masses (involving further LECs), the presence of which will effectively be taken into account in the fit ansatz of Eq.~(\ref{eq:pion_fit}).
After inserting the $\rm{SU}(3)$ ChPT prediction for the pion decay constant at NLO~\cite{GASSER1985465} 
\begin{align}
f_{\pi} &= f_{0}\left[ 1 - 2\left(\frac{M_{\pi}}{4\pi f_{0}}\right)^{2}\log{\left(\frac{M_{\pi}}{4\pi f_{0}}\right)^{2} }\right. \nonumber \\[6pt]
&\left. -\left(\frac{M_{K}}{4\pi f_{0}}\right)^{2}\log{ \left(\frac{M_{K}}{4\pi f_{0}}\right)^{2}}\right] + \mathcal{O}\left(M_{\pi}^{2}, M_{K}^{2}\right)~,
\end{align}
we obtain for the ratio $R_{\pi}$ the expression
\begin{align}
\label{eq:R_pi_chir}
R_{\pi} = 4e^{2}C - 3e^{2}\left(\frac{M_{\pi}}{4\pi f_{0}}\right)^{2}\log{\left(\frac{M_{\pi}}{4\pi f_{0}}\right)^{2} } + \mathcal{O}\left(M_{\pi}^{2}, M_{K}^{2}\right)~.    
\end{align}
Notice that the chiral prediction for the ratio $R_{\pi}$ is not affected, even in the $\rm{SU}(3)$ effective theory, by logarithmic corrections in the kaon mass, at NLO. \\

Inspired by the ChPT prediction of Eq.~(\ref{eq:R_pi_chir}), and by the non-relativistic expansion of Ref.~\cite{Davoudi:2014qua}, we extrapolate the lattice data towards the physical pion mass and towards the continuum and infinite volume limit, employing the following ansatz for the ratio $R_{\pi}$
\begin{align}
R_{\pi}^{\textrm{sub.}}(\xi_{\pi}, a, L) &=  4e^{2}C -3e^{2} \xi_{\pi}\log{\xi_{\pi}} + e^{2}A_{1} \xi_{\pi}   \nonumber \\[6pt]
	 				    &+ e^{2}A_2 \xi_{\pi}^{2} +  e^{2}D\,a^2 +  e^{2}D_m \xi_{\pi} a^2   \nonumber \\[6pt]
	 				    &+ e^{2}K \frac{(4\pi)^{2}\xi_{\pi}}{3M_{\pi}L^3} \langle r^2 \rangle_{\pi^+}\left( 1 + \frac{F_{4}}{M_{\pi}L}\right)  \nonumber \\[6pt]
	 				    &+ e^{2}F_{a}\frac{\xi_{\pi}}{M_{\pi}}\frac{a^{2}}{L^{3}}  ~,
    \label{eq:pion_fit}
\end{align}
where $R_{\pi}^{\textrm{sub.}}$ is the $R_{\pi}$ ratio after the subtraction of the universal FSEs using Eq.~(\ref{eq:universal_FSE}), and $\xi_{\pi} \equiv M_{\pi}^{2}/(4\pi f_{\pi})^{2}$ after applying the CDH corrections to $M_{\pi}$ and $f_{\pi}$. In the previous expression $C, A_{1}, A_{2}, D, D_{m}, K, F_{a}$ and $F_{4}$ are treated as free fitting parameters. In particular $C$ and $A_{1}$ parameterize the ChPT expansion for $R_{\pi}$ up to NLO, $A_{2}$ is an effective  LEC at NNLO, while $D$ and $D_{m}$ take into account discretization effects. The constant $K$ parametrizes deviations from the non-relativistic prediction ($K=1$) of Eq.~(\ref{eq:pion_savage}) for the SD FSEs, the term proportional to $a^{2}/L^{3}$ corresponds to a FSE due to an heavy intermediate state of mass $\propto 1/a$~\cite{Tantalo:2016vxk}, and finally $F_{4}$ parametrizes higher order $\mathcal{O}(1/L^{4})$ FSEs. \\

\begin{figure*}
\begin{center}
\includegraphics[scale=0.65]{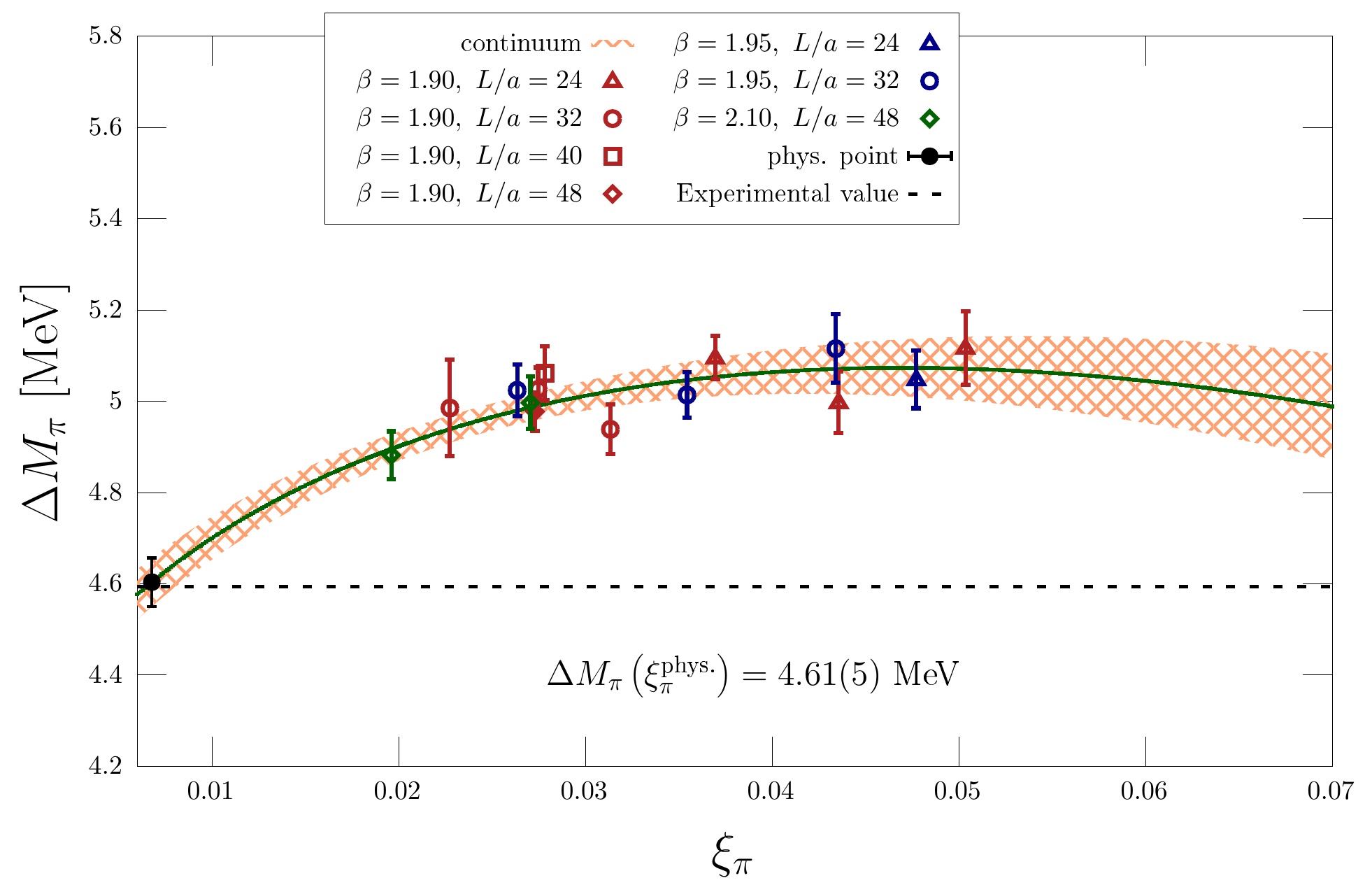}
\caption{\it\small Our results for $\Delta M_{\pi}$  as a function of the parameter $\xi_{\pi}$. The data points refer to our lattice estimate after the subtraction of both universal and SD FSEs. The solid lines represents, for each lattice spacing, the central value of the fitted curve in Eq.~(\ref{eq:pion_fit}) obtained setting $A_{2}=D=D_{m}=F_{a}=F_{4}=0$, and in the thermodynamic limit $L\to\infty$. The orange band shows the statistical uncertainty  on $\Delta M_{\pi}$ after the continuum and infinite volume extrapolation. Finally, the black point corresponds to our determination at the physical point $\xi_{\pi}^{\textrm{phys.}} \simeq 0.00678~ \textrm{MeV}$.}
\label{fig:fit_dim}
\end{center}
\end{figure*}

In Fig.~(\ref{fig:fit_dim}), we show the result of the extrapolation obtained using the ansatz of Eq.~(\ref{eq:pion_fit}), setting $A_{2} = D = D_m = F_{a}= F_{4} = 0$, which corresponds to our preferred fit. The quantity $\Delta M_{\pi}$, which at the physical point gives the pion mass splitting, is defined as 
\begin{align}
\Delta M_{\pi} \equiv R_{\pi}  \frac{\left(f_{\pi}^{\rm{phys.}}\right)^{2}}{2M_{\pi}^{\rm{phys.}}}~,
\end{align}
with $M_{\pi}^{\rm{phys.}} = 134.977~{\rm MeV}$ and $f_{\pi}^{\rm{phys.}} = 130.4~{\rm MeV}$. 
The reduced $\chi^{2}$ of the fit is $\chi^{2}/d.o.f. \sim 1.8$ with $14$ measures and 4 parameters. Notice the remarkable smallness of $\mathcal{O}(a^{2})$ effects in our data. The inclusion of the fit parameter $D$ to describe $\mathcal{O}(a^{2})$ lattice artifacts does not improve the description of the data: the resulting $\chi^{2}/d.o.f.$ is $1.9$ and the parameter $D$ turns out to be consistent with zero within errors ($D=0.04(10)$). To estimate systematic errors, we perform a total of $24$ fits, differing on whether the $A_{2}$, the $D$, and the $D_{m}$ fit parameters are included or not, and on the form of the SD FSEs for which we either include $K$ or $F_{a}$ as a free fit parameter (in this last case setting $K=1$), or include the additional term $\propto F_{4}/L^{4}$ on top of the non-relativistic prediction $K=1, F_{a}=0$. The fit results are combined using the Akaike Information Criterion (AIC)~\cite{AIC}, in which, to each fit, it is assigned a weight $w_{i} \propto \exp{-(\chi^{2}+2n_{pars})/2}$.
Mean values and standard errors are then computed using~\cite{Carrasco:2014cwa,alexandrou2021quark, Jay:2020jkz}
\begin{align}
\bar{x} = \sum_{i}~w_{i}\, \bar{x}_{i}~,\quad \sigma^{2} = \sum_{i}~w_{i}\,\left(\sigma_{i}^{2} + (\bar{x}_{i} - \bar{x})^{2}\right)~, 
\end{align}
where $(\bar{x}_{i}, \sigma_{i})$ is the mean value and the standard error obtained in the $i-$th fit.
Our final result for the pion mass splitting is
\begin{align}
\label{eq:final_result_pion}
M_{\pi^{+}}- M_{\pi^{0}} &= 4.622~(64)_{stat.}(70)_{syst.}~{\rm MeV} \nonumber \\[8pt]
&= 4.622~(95)~{\rm MeV}~,
\end{align}
which agrees very well with the experimental determination $
\left[ M_{\pi^{+}}- M_{\pi^{0}}\right]^{\emph{exp.}} \,=\, 4.5936~(5)~ \textrm{MeV}$, 
and with the result of a recent lattice determination~\cite{Feng:2021zek} $M_{\pi^+} - M_{\pi^0} = 4.534(42)(43)~{\rm MeV}$,  in which the disconnected contribution has been computed as well.  \\

\section{Conclusions}
\label{sec:conclusions}
We have presented an analysis of the $\mathcal{O}(\alpha_{em})$ mass splitting $M_{\pi^{+}}-M_{\pi^{0}}$ between the charged and neutral pion, including the calculation of the disconnected diagram. We made use of the gauge configurations generated by the Extended Twisted Mass Collaboration with $N_{f}=2+1+1$ dynamical quark flavours. The gauge ensembles considered corresponds to three different values of the lattice spacing $a\simeq 0.062,0.082$ and $0.089~{\rm fm}$, pion masses in the range $M_{\pi} \simeq 250-450~{\rm MeV}$, while the strange and charm quark masses are set in all ensembles to their physical value. We showed that a good accuracy in the determination of the disconnected diagram can be achieved by working in the \textit{rotated twisted mass} (RTM) scheme, which have been shown to be particularly convenient for the evaluation of some QCD$+$QED mesonic observables based on the RM123 approach. We also developed a new numerical technique, tailored to quark-line disconnected diagrams, which does not rely on a stochastic representation of the photon propagator and that produced a further reduction of the statistical noise by more than one order of magnitude. After extrapolating to the continuum and infinite volume limit, and at the physical point, we obtain a value for $M_{\pi^{+}}-M_{\pi^{0}}$ which perfectly agrees with the experimental result. With respect to our previous determination~\cite{Giusti:2017dmp}, we were able to reduce the uncertainty on the pion mass splitting by a factor $\sim 3$, thanks to the use of the dimensionless ratio $R_{\pi}$, and its ChPT-based extrapolation analysis, and of larger lattice volumes.
\section{Acknowledgement}
We thank C. Tarantino for useful discussions, and all members of ETMC for the most enjoyable collaboration. We acknowledge CINECA for the provision
of CPU time under the specific initiative INFN-LQCD123 and IscrB\_S-EPIC. F.S. G.G and S.S. are supported by
the Italian Ministry of University and Research (MIUR) under grant PRIN20172LNEEZ. F.S. and G.G are supported by INFN under GRANT73/CALAT.

\appendix
\section{Numerical methods for the disconnected diagram}
\label{sec:num_methods}
The disconnected contribution to the pion mass difference $M_{\pi^{+}}- M_{\pi^{0}}$, in the RTM basis, is encoded in the following Wick-contracted Green function of the isospin symmetric theory
\beq
\label{eq:def_G4}
G^{(4)}_{\mu\nu}(t-t', y, y') &=& \frac{1}{L^{3}}\sum_{\vec{x},\vec{x}'}\langle 0 | T \left\{ \wick[arrows={W->-,W-<-},below]{\c1 \phi_{\pi^{\prime -}}\c2(\vec{x},t) \c1 J_{\mu}^{ib}\c2 (y)}~\cdot\right . \nonumber \\[8pt]
&\cdot& \left. \wick[arrows={W->-,W-<-},below]{ \c1 J_{\nu}^{ib}\c2 (y')\c1 \phi_{\pi^{\prime +}}^{\dag}\c2 (\vec{x}',t')}\right\}|0\rangle~,
\eeq
Eq.~(\ref{eq:def_G4}) can be in turn written as
\begin{align}
G^{(4)}_{\mu\nu}(t-t',y,y')  = 2e^2(\Delta q)^{2}\cdot\langle
G_{\mu}^{(2)}(t,y)G_{\nu}^{(2)}(t',y')\rangle_{U}\, ,
\end{align}
where $\langle\, .\, \rangle_{U}$ denotes the average over the $\rm{SU}(3)$ QCD gauge field, while
\begin{align}
G_{\mu}^{(2)}(t,y) & =   \frac{1}{\sqrt{L^{3}}}\sum_{\vec{x}} \tr \left\{ S_{\ell}^{+}( (\vec{x},t), y)\gamma_{\mu}S^{-}_{\ell}(y, (\vec{x},t))\gamma_{5}\right\}~, 
\label{eq:G4}
\end{align}
and $\tr$ is meant over color and spin indices. In the
previous equation, $S_\ell^\pm$ is the propagator of the two ``flavour''
components corresponding to Wilson parameters $r = \pm 1$ of the light 
quark isodoublet field $\psi'_\ell$ entering the RTM action in Eq.~(\ref{eq:RTM}). \\

As illustrated in Sec.~[\ref{sec:RM123}], the disconnected contribution to $M_{\pi^{+}}- M_{\pi^{0}}$, can be extracted from the large time behavior of $\delta C'^{disc}_{\pi}(t)/C_{\pi\pi}^{\textrm{isoQCD}}(t)$, where $C_{\pi\pi}^{\textrm{isoQCD}}(t)$ is the TM charged pion correlator
\begin{align}
C_{\pi\pi}^{\textrm{isoQCD}}(t) = \frac{1}{\sqrt{L^{3}}}\sum_{\vec{x}}\langle 0 | T \left \{  \phi_{\pi^{ +}}(\vec{x},t)\,\phi_{\pi^{ +}}^{\dag}(0)\right\} | 0  \rangle~,
\end{align}
and 
\begin{align}
\delta C'^{disc.}_{\pi}(t'-t) = \frac{1}{2e^{2}(\Delta q)} \sum_{y,y'} G^{(4)}_{\mu\nu}(t,t',y,y')\Delta_{\mu\nu}(y,y')~,
\label{eq:disc_def}
\end{align}
As usual, the QED corrections at order $\mathcal{O}(\alpha_{em})$ require the computation of the integrals over the two ends of the photon propagator (sum over $y$ and $y'$ in Eq.~(\ref{eq:disc_def})). The explicit summation is, however, prohibitively costly on a large four dimensional lattice, as it scales like the square of the lattice volume $V=L^{3}\cdot T$. To cope with this issue, in Ref.~\cite{Giusti:2017dmp} a stochastic technique has been adopted to evaluate all connected diagrams arising at order $\mathcal{O}(\alpha_{em})$, and in this work we adopt the same strategy to compute the exchange diagram. The idea of the stochastic approach is to exploit the definition of the photon propagator in terms of the expectation value of the time ordered product of photon fields, i.e.
\beq
\Delta_{\mu\nu}(y-y') = \langle A_{\mu}(y) A_{\nu}(y') \rangle_{A} ~.
\eeq 
One can then sample each mode of the photon field $A_{\mu}(y)$ from the local probability distribution in momentum space
\beq
P\left[ \vec{A}(k)\right] d\vec{A}(k) \propto \exp{\left(-A_{\mu}(k)\Delta^{-1}_{\mu\nu}(k)A_{\nu}(k)\right)}~, 
\eeq
where in Feynman gauge $\Delta^{-1}_{\mu\nu}(k)= \delta_{\mu\nu}/\tilde{k}^{2}$, with
\begin{align}
\tilde{k}^{\nu} = 2\sin{\left( \frac{k^{\nu}}{2} \right)}~.   
\end{align}
In our $QED_L$ setup all the ``spatial zero modes'' $A_{\mu}(k^{0}, \vec{k} = 0)$ are removed. By drawing a sample $\{A_{\mu}^{i}\}_{i=1,\ldots, n}$, the photon propagator can be estimated as
\beq
\Delta_{\mu\nu}(y-y')  \simeq \frac{1}{n}\sum_{i=1}^{n} A_{\mu}^{i}(y)A_{\nu}^{i}(y')~,
\eeq
and the estimate becomes an exact equality in the infinite $n$ limit. 
In this way, any observable of the form
\beq
\mathcal{O} = \sum_{y,y'}\,\langle O_{1}^{\mu}(y)\cdot\mathcal{O}_{2}^{\nu}(y')\rangle_{U}\Delta_{\mu\nu}(y,y')~,
\eeq
can be split into two summations, each scaling as the lattice volume $V$, using
\beq
&\mathcal{O}&  =  \lim_{n\to\infty}\frac{1}{n}\sum_{i}\,\langle\mathcal{O}_{1}^{A^{i}}\cdot\mathcal{O}_{2}^{A^{i}}\rangle_{U}~, \nonumber \\[6pt]
&\mathcal{O}_{j}^{A^{i}} & = \sum_{y}\, \mathcal{O}_{j}^{\mu}(y)A_{\mu}^{i}(y)\, ,\quad j=1,2~.
\eeq
We refer to the Appendix A of Ref.~\cite{Giusti:2017dmp} for a detailed explanation on how this strategy can be implemented to efficiently evaluate the exchange diagram.\\

However, despite the stochastic technique proves itself useful in many occasions,  the evaluation of the disconnected diagram turns out to be very noisy, requiring a too large value of $n$ to get a clear signal.  In this work we thus adopt a different strategy, specific to quark-line disconnected diagrams, that does not rely on a stochastic representation of the photon propagator, and allows to compute $\delta C'^{disc.}_{\pi}$ in $\mathcal{O}\left( V\log V\right)$ time. \\

Starting from Eq.~(\ref{eq:G4}), we define
\beq
\tilde{B}_{\nu}(t,y')  \defs \sum_{y} G^{(2)}_{\mu}(t,y)\Delta_{\mu\nu}(y-y') ~,
\eeq
which is nothing but the convolution $G^{(2)}_{\mu}(t)\ast \Delta_{\mu\nu}$. In Fourier space we thus have
\beq
\tilde{B}_{\nu}(t,k) = G_{\mu}^{(2)}(t,k)\Delta_{\mu\nu}(k)= G_{\nu}^{(2)}(t,k)/ \tilde{k}^{2}~,
\eeq
where on a $L^{3}\times T$ torus and within the $QED_{L}$ theory we considered, the set of allowed momenta $k$ is given by
\beq 
&k&  = 2\pi\left(\frac{n_{0}}{T}, \frac{n_{1}}{L}, \frac{n_{2}}{L}, \frac{n_{3}}{L}\right)~, \\[6pt]
&n_{0}&  \in  \left\{0,\ldots, T-1\right\}\, , \\[6pt] 
&n_{i=1,2,3}&  \in  \left\{1,\ldots, L-1\right\}~.
\eeq
The quantity $G_{\mu}^{(2)}(t,k)$ can be computed from $G_{\mu}^{(2)}(t,x)$ using  Fast Fourier Transform (FFT) methods in $\mathcal{O}(V\log{V})$ time. Moreover, transforming back from $\tilde{B}(t,k)$ to $\tilde{B}(t,y')$ via
\beq
\tilde{B}_{\nu}(t,y') = \sum_{k} e^{-iky'} \tilde{B}_{\nu}(t,k)\, ,
\eeq
can be done within a similar machine time.
The disconnected diagram $\delta C'^{disc.}_{\pi}(t'-t)$ is then obtained from the  product between the "bubble diagram"  $G^{(2)}_{\mu}(t',y')$ and $\tilde{B}_{\mu}(t,y')$ via
\begin{align}
\delta C'^{disc.}_{\pi}(t'- t) = \sum_{y'} \tilde{B}_{\nu}(t,y')G^{(2)}_{\nu}(t',y')~, 
\end{align}
which requires $\mathcal{O}(V)$ operations, and the overall complexity of the computation is thus of order $\mathcal{O}(V\log{V})$. \\

Concerning the bubble diagram $G^{(2)}_{\mu}(t,y)$, for each gauge configuration its value can be estimated through a single inversion of the lattice Dirac operator as in the case of the pion correlator of the isospin symmetric theory. As is standard practice, the inversion is performed stochastically using a certain number $N_{s}(t)$ of random sources. This induces, as usual, an uncertainty on $G^{(2)}_{\mu}(t,y)$ that scales as $(N_{s}(t))^{-1/2}$. The noise induced by the use of the stochastic method on $\delta C'^{disc.}_{\pi}(\Delta t = t'-t)$ at a given $\Delta t$, depends instead on two factors. The first one is clearly the total number $N(\Delta t)$ of stochastic sources separated by a time distance $\Delta t$, i.e.
\begin{align}
N(\Delta t) = {\displaystyle \sum_{t=0}^{T-1}} N_{s}(t) \cdot N_{s}((t+\Delta t)\!\!\!\mod{T})~.     
\end{align}
Indeed, defining $\delta C'^{disc.}_{\pi}(\Delta t, t_{i}, t'_{j})$ to be the estimate of $\delta C'^{disc.}_{\pi}(\Delta t)$ obtained considering only the $i-$th random source at time $t$ and the $j-$th random source at time $t'= (t+\Delta t )\!\!\!\mod{T}$, the value of $\delta C'^{disc.}_{\pi}(\Delta t)$ is estimated through
\begin{align}
\label{eq:ave_disc}
\delta C'^{disc.}_{\pi}(\Delta t) \sim \frac{1}{N(\Delta t)} \sum_{t=0}^{T-1}\sum_{i=1}^{N_{s}(t)}\sum_{j=1}^{N_{s}(t')} \delta C'^{disc.}_{\pi}(\Delta t, t_{i}, t'_{j})~.    
\end{align}
However, the statistical uncertainty on $\delta C'^{disc.}_{\pi}(\Delta t )$ does not scale exactly as $\left( N(\Delta t)\right)^{-1/2}$ since a given random source appears in general many times in the sum of Eq.~(\ref{eq:ave_disc}), inducing a correlation between the various terms. Thus, understanding how the uncertainty on $\delta C'^{disc.}_{\pi}(\Delta t)$ behaves depending on the choice of the $N_{s}(t)$ is a non-trivial problem. Given a fixed number of sources $N_{s}^{tot} = \sum_{t=0}^{T-1} N_{s}(t)$, one would like to find the optimal way of distributing the $N_{s}^{tot}$ sources along the time direction, in such a way that the signal of $\delta C'^{disc.}_{\pi}(\Delta t)$ is optimized in a given interval $\Delta t \in [t_{min}, t_{max}]$ where one expects the ground-state to be dominant. However, after performing preliminary tests using $N_{s}^{tot} = T$, we found that the uniform distribution $N_{s}(t) = 1$, produces the smallest errors among various choices of the distribution of the random sources. The results presented in Sec.~[\ref{sec:numerical_results}] have been obtained employing $N_{s}(t)=1$ for all simulation points.\\

\begin{figure*}
    \centering
    \includegraphics[scale=0.74]{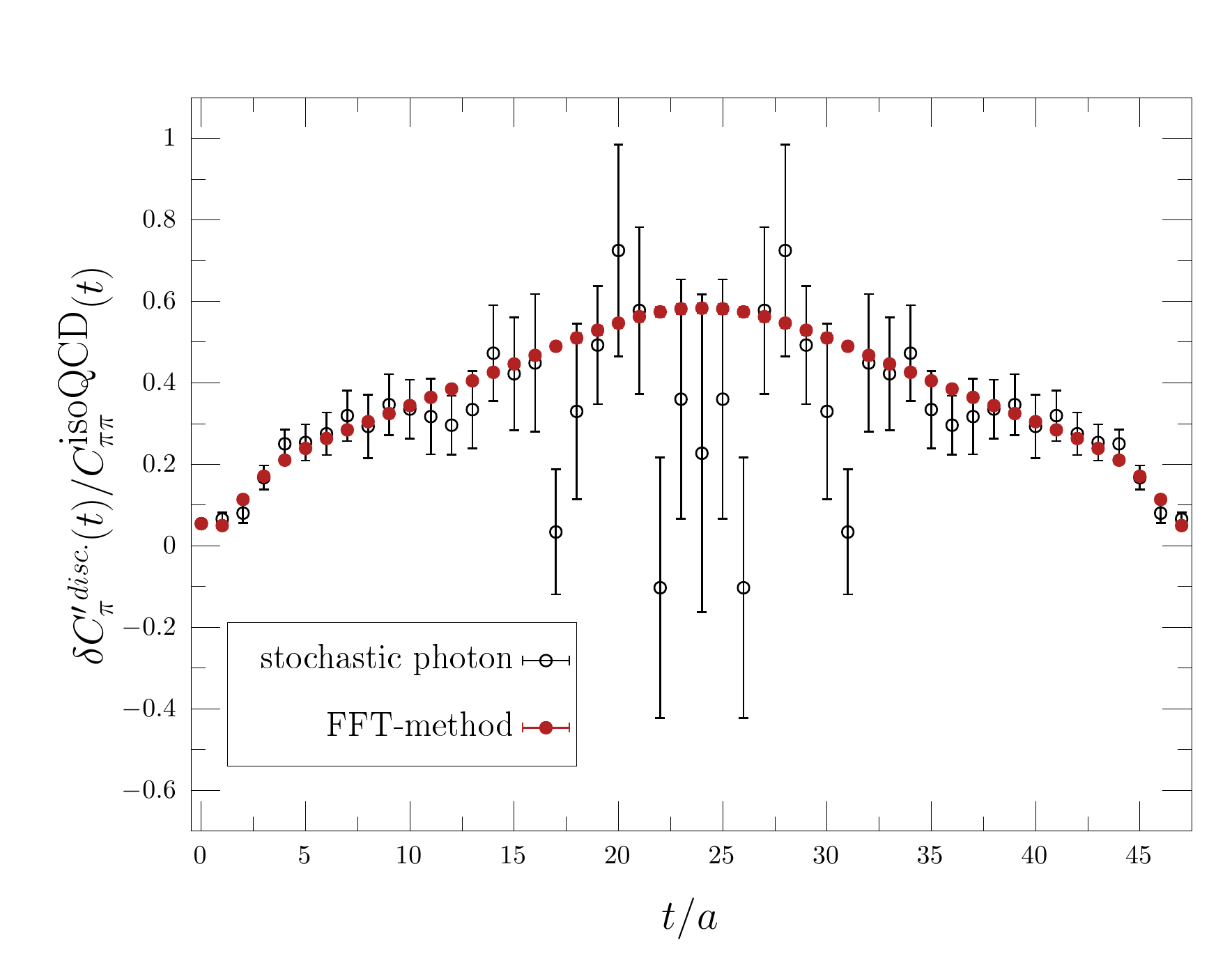}
    \caption{Ratio of correlators $\delta C'^{disc.}_{\pi}(t)/C_{\pi\pi}^{\textrm{isoQCD}}(t)$ computed using the FFT-based method (solid points), and the stochastic one in which the photon field $A_{\mu}$ is sampled (empty points). The two datasets correspond to simulations on a $24^{3}\times 48$ lattice at $\beta=1.90$ having a similar numerical cost. }
    \label{fig:comparison_stochastic_exact}
\end{figure*}

Finally, we show in Fig.~(\ref{fig:comparison_stochastic_exact}) a comparison between the ratio of correlators $\delta C'^{disc.}_{\pi}(t)/C_{\pi\pi}^{\textrm{isoQCD}}(t)$ computed using both the FFT-based and the stochastic method. The comparison is performed at approximately equal total machine time, on a $24^{3}\times 48$ lattice at $\beta=1.90$ and with a pion mass $M_{\pi} \sim 320~ \textrm{MeV}$.  As it can be seen from the figure the improvement is dramatic, with the precision on $\delta C'^{disc.}_{\pi}(t)/C_{\pi\pi}^{\textrm{isoQCD}}(t)$ improving by a factor $10-20$. 

\begin{thebibliography}{99}

\bibitem{Aoki:2021kgd}
Y.~Aoki, T.~Blum, G.~Colangelo, S.~Collins, M.~Della Morte, P.~Dimopoulos, S.~D\"urr, X.~Feng, H.~Fukaya and M.~Golterman, et al.
{\it FLAG Review 2021}. e-Print: 2111.09849 [hep-lat].



\bibitem{Borsanyi:2014jba}
S.~Borsanyi, S.~Durr, Z.~Fodor, C.~Hoelbling, S.~D.~Katz, S.~Krieg, L.~Lellouch, T.~Lippert, A.~Portelli and K.~K.~Szabo, \textit{et al.}
{\it Ab initio calculation of the neutron-proton mass difference},
Science 347 (2015), 1452-1455. e-Print: 1406.4088 [hep-lat].

\bibitem{Boyle:2017gzv}
P.~Boyle, V.~G\"ulpers, J.~Harrison, A.~J\"uttner, C.~Lehner, A.~Portelli and C.~T.~Sachrajda,
{\it Isospin breaking corrections to meson masses and the hadronic vacuum polarization: a comparative study},
J. High Energ. Phys. 2017, 153 (2017). e-Print: 1706.05293 [hep-lat].

\bibitem{Hansen:2018zre}
M.~Hansen, B.~Lucini, A.~Patella and N.~Tantalo,
{\it Gauge invariant determination of charged hadron masses},
J. High Energ. Phys. 2018, 146 (2018). e-Print: 1802.05474 [hep-lat].

\bibitem{IBs}
RM123 collaboration, G. M. de Divitiis et al.,
{\it Isospin breaking effects due to the up-down mass difference in lattice QCD}, 
J. High Energ. Phys. 2012, 124 (2012). e-Print: 1110.6294 [hep-lat].


\bibitem{deDivitiis:2013xla}
RM123 collaboration, G. M. de Divitiis et al., 
{\it Leading isospin breaking effects on the lattice}, 
Phys.Rev.D 87 (2013) 11, 114505. e-Print: 1303.4896 [hep-lat].

\bibitem{lep1}
RM123 collaboration, D.~Giusti et al., 
{\it First lattice calculation of the QED corrections to leptonic decay rates},
Phys. Rev. Lett. 120 (2018) 7, 072001. e-Print: 1711.06537 [hep-lat].


\bibitem{lep2}
RM123 collaboration M.~Di Carlo et al., 
{\it Light-meson leptonic decay rates in lattice QCD$+$QED},
Phys. Rev. D 100 (2019) 3, 034514. e-Print: 1904.08731 [hep-lat].

\bibitem{lep3}
RM123 collaboration A.~Desiderio et al., 
{\it First lattice calculation of radiative leptonic decay rates of pseudoscalar mesons},
Phys. Rev. D 103 (2021) 1, 014502. e-Print: 2006.05358 [hep-lat].

\bibitem{gm1}
RM123 collaboration D.~Giusti et al., 
{\it Strange and charm HVP contributions to the muon $(g - 2)$ including QED corrections with twisted-mass fermions},
J. High Energ. Phys. 2017, 157 (2017). e-Print: 1707.03019 [hep-lat].

\bibitem{gm2}
RM123 collaboration D.~Giusti et al., 
{\it Electromagnetic and strong isospin-breaking corrections to the muon $g - 2$ from Lattice QCD$+$QED},
Phys. Rev. D 99 (2019) 11, 114502. e-Print: 1901.10462 [hep-lat].


\bibitem{Giusti:2017dmp}
RM123 collaboration, D. Giusti et al.,
{\it Leading isospin-breaking corrections to pion, kaon and charmed-meson masses with Twisted-Mass fermions},
Phys.Rev.D 95 (2017) 11, 114504. e-Print: 1704.06561 [hep-lat].

\bibitem{PhysRev.183.1245}
R.~F.~Dashen,
{\it Chiral $\mathrm{SU}(3)\ensuremath{\bigotimes}\mathrm{SU}(3)$ as a symmetry of the strong interactions},
Phys. Rev. 183 (1969) 5, 1245-1260.

\bibitem{Fr:2021}
R.~Frezzotti, G.~Gagliardi, V.~Lubicz, F.~Sanfilippo and S.~Simula,
{\it Rotated twisted-mass: a convenient regularization scheme for isospin breaking QCD and QED lattice calculations},
Eur. Phys. J. A 57 (2021) 9, 282. e-Print: 2106.07107 [hep-lat].  
  

\bibitem{Baron:2010bv}
ETM collaboration, R.~Baron et al.,
{\it Light hadrons from lattice QCD with light $(u,d)$, strange and charm dynamical quarks}, 
J. High Energ. Phys. 2010, 111 (2010). e-Print: 1004.5284 [hep-lat].


\bibitem{Carrasco:2014cwa}
ETM collaboration, N. Carrasco et al.,
{\it Up, down, strange and charm quark masses with N$_f$ = 2+1+1 twisted mass lattice QCD},  
Nucl.Phys.B 887 (2014), 19-68. e-Print: 1403.4504 [hep-lat].

\bibitem{pdg}
P.~A.~Zyla et al., [Particle Data Group],
{\it Review of Particle Physics},
PTEP  2020 (2020) no.8, 083C01.

\bibitem{Frezzotti:2003ni}
R. Frezzotti and G.C. Rossi, 
{\it Chirally improving Wilson fermions 1. O(a) improvement}, 
J. High Energ. Phys. 2004, 08 (2004). e-Print: hep-lat/0306014 [hep-lat].


\bibitem{Frezzotti:2004wz}
R.~Frezzotti and G.~C.~Rossi,
{\it Chirally improving Wilson fermions. II. Four-quark operators}, 
J. High Energ. Phys. 2004, 10 (2004). e-Print: hep-lat/0407002 [hep-lat].


\bibitem{Colangelo:2005gd}
G.~Colangelo, S.~Durr and C.~Haefeli,
{\it Finite volume effects for meson masses and decay constants},
Nucl. Phys. B \textbf{721} (2005), 136-174. e-Print: hep-lat/0503014  [hep-lat].


\bibitem{Davoudi:2014qua}
Z.~Davoudi and M.~J.~Savage,
{\it Finite-Volume Electromagnetic Corrections to the Masses of Mesons, Baryons and Nuclei},
Phys. Rev. D 90 (2014) 5, 054503. e-Print: 1402.6741 [hep-lat].


 \bibitem{Hayakawa:2008an}
M.~Hayakawa and S.~Uno,
{\it QED in finite volume and finite size scaling effect on electromagnetic properties of hadrons},
Prog. Theor. Phys. 120 (2008) 3, 413-441. e-Print: 0804.2044 [hep-ph].
 
 \bibitem{GASSER1985465}
J.~Gasser and H.~Leutwyler, {\it Chiral perturbation theory: Expansions in the
  mass of the strange quark}, {\em Nuclear Physics B}, vol.~250, no.~1,
  pp.~465--516, 1985.
  
\bibitem{Tantalo:2016vxk}
N.~Tantalo, V.~Lubicz, G.~Martinelli, C.~T.~Sachrajda, F.~Sanfilippo and S.~Simula,
{\it Electromagnetic corrections to leptonic decay rates of charged pseudoscalar mesons: finite-volume effects}, PoS LATTICE2016 (2016), 300. e-Print: 1612.00199 [hep-lat].

\bibitem{AIC} 
H. Akaike,
{\it A new look at the statistical model identification}, IEEE Transactions on Automatic Control 19, 716 (1974).

\bibitem{alexandrou2021quark}
ETM collaboration, C.~Alexandrou et al,
{\it Quark masses using twisted-mass fermion gauge ensembles},
Phys. Rev. D 04 (2021) 7, 074515. e-Print: 2104.13408 [hep-lat].  

\bibitem{Jay:2020jkz}
W.~I.~Jay and E.~T.~Neil,
{\it Bayesian model averaging for analysis of lattice field theory results},
Phys. Rev. D 103 (2021), 114502. e-Print: 2008.01069 [stat.ME].

\bibitem{Feng:2021zek}
X.~Feng, L.~Jin and M.~J.~Riberdy,
{\it Lattice QCD calculation of the pion mass splitting}, e-Print: 2108.05311 [hep-lat].








\end{thebibliography}
\end{document}